# HOW COBIT CAN COMPLEMENT ITIL TO ACHIEVE BIT

[1,] Narges Zeinolabedin *, [2,] Soroush Afiati Mehrvarz
[3,] Neda Rahbar

[1] Department of ITM, Islamic Azad University, Electronic Branch, Tehran, Iran
[2] Department of ITM, Islamic Azad University, Electronic Branch, Tehran, Iran
[3] Department of Computer, Islamic Azad University, Broujerd Branch, Broujerd, Iran

*Abstract:* *Strategic alignment is a conviction that is considered extremely important in understanding how organizations can apply their arrangement of information technology (IT) into substantial boosts in achievement. To attain alignment advantage, Information Technology Infrastructure Library (ITIL) prepares a framework of best practice approch for IT Service Management in all countries and Control Objectives for Information and Related Technology (COBIT) is an IT governance framework and aiding toolset that permits managers to stretch the gap between control prerequisites, technical matters and business risks. The purpose of this paper is to recognize how COBIT can complement ITIL to attain Business-IT Alignment.*

*Keywords:* *Information Technology Infrastructure Library (ITIL); Control Objectives for Information and Related Technology (COBIT); Business-IT alignment (BITA).*

## 1 INTRODUCTION:

Information technology (IT) is changing the way companies organize their business processes, communicate with their customers and potential customers, and deliver their services (Silvius , 2011). A key factor for a successful company is an effective and efficient alignment of the way IT supports business Strategies and processes (Silvius, Waal, & Smit, 2009). The necessity and desirability of aligning business needs and IT capabilities has been examined in numerous articles and its importance is well recognized (Silvius, Waal, & Smit, 2009). The annual survey of top management concerns by the Society for Information Management (www.simnet.org) ranked 'IT and Business alignment' as the no. 1 concern in five of the last six years (Society of Information Management, 2003, 2004, 2005, 2006, 2007, 2008) (Silvius, Waal, & Smit, 2009). In the year that it did not make the top spot, alignment ranked as the no. 2 concern. The Alignment between business needs and IT abilities is therefore yet a notable area of consternation (Silvius, et al., 2011) .The purpose of this study is to donate an awareness into the attainment of alignment between IT and business targets by examining how organizations move toward Information Technology Service Management (ITSM). One recent development in ITSM is the universal adoption of ITIL in order to supply the best possible delivery of IT services (Kashanchi & Toland, 2006).

In addition, COBIT is an IT governance framework and supporting toolset that allows managers to bridge the gap between control requirements, technical issues and business risks (ISACA, 2013). COBIT enables clear policy development and good practice for IT control throughout organizations (ISACA, 2013). COBIT emphasizes regulatory compliance, helps organizations to increase the value attained from IT, enables alignment and simplifies implementation of the COBIT framework (ISACA, 2013).

This research concentrates on the applying of SAM model (Henderson & Venkatraman, 1999) to recognize ITIL and COBIT roles the BITA in Iranian organizations.

## 2 LITERATURE REVIEW

To evolve a hypothetical framework for assessing how COBIT can enhance ITIL to attain BITA, several succession of related literature were appraised. This included of studying preceding and recent research literature to associate the nature of research that has been done consequently far and what is even to be investigated, in the time to come.

### 2.1 Business-IT alignment:

Business-IT alignment is the highly desired state in which an organization can effectively use IT to achieve its business objectives (Rahbar, Zeinolabedin, & Afiati, 2013). This expression particularly encompasses the efforts of the IT and business professionals to bridge the gap prevalent among the stakeholders of the organization, owing to differences in objectives, culture and incentives, including a reciprocal unawareness of one another (Rahbar, Zeinolabedin, & Afiati, 2013).

The notion of strategic alignment is more than two decades old (Papp, 2001).The original alignment model was a largely theoretical construct that studied only a single industry but has been adapted for use by virtually any industry looking to integrate their business strategies with their information technology strategies (Papp, 2001).

### 2.2 ITIL:

One possible way to achieve alignment is for IT organizations to transform themselves into service providers (Rahbar, Zeinolabedin, & Afiati, 2013). Being a service provider means using IT as a solution to business problems and running the IT department as a business function (Rahbar, Zeinolabedin, & Afiati, 2013). It also means providing a new competitive strategy (Rahbar, Zeinolabedin, & Afiati, 2013). This is because the focus of the companies moves toward customers and providing high quality products and services at low cost to satisfy their demands (Rahbar, Zeinolabedin, & Afiati, 2013). In order to be a potent service provider organizations are demanded to have noble quality ITSM (Kashanchi & Toland, 2006). ITSM is "concerned with delivering and supporting IT services that are appropriate to the business requirements of the organization" (Rahbar, Zeinolabedin, & Afiati, 2013). ITSM applies the best practice ITIL approach to enhance delivery and support of IT services (Rahbar, Zeinolabedin, & Afiati, 2013). Besides, ITIL will enable organizations to ameliorate their IT service management (Kashanchi & Toland, 2006).

Information Technology Infrastructure Library (ITIL) has introduced with Office of Government Commerce (OGC) (Rahbar, Zeinolabedin, & Afiati, 2013). It has three versions: ITIL Version 1, ITIL Version 2 and recently ITIL Version 3(Rahbar, Zeinolabedin, & Afiati, 2013). After ten years use of ITIL V2, ITIL V3 was introduced in 2007 by OGC (Rahbar, Zeinolabedin, & Afiati, 2013). The context of this publication of the ITIL is the ITIL framework as a source of good practice in service management (Rahbar, Zeinolabedin, & Afiati, 2013).

For better understanding of ITIL, here some difference of ITIL V2 and V3 will discussed (Rahbar, Zeinolabedin, & Afiati, 2013). "ITIL Version 2 deals primarily with aligning IT to the business. But ITIL V3 will enable organizations to move from alignment of IT with the business to the integration of IT with the business" (Rahbar, Zeinolabedin, & Afiati, 2013). The ITIL Version 3 Library has the following components (Rahbar, Zeinolabedin, & Afiati, 2013):

• **The ITIL Core**:

There are five volumes as best practice guidance applicable which covers all types of organizations who provide services to a business (Rahbar, Zeinolabedin, & Afiati, 2013). So, the construction of the basis is in the shape of a lifecycle and it is iterative and multidimensional.

• **The ITIL Complementary Guidance**:

A complementary set of publications which are useful guides for industry sectors, organization types, operating models, and technology architectures (Rahbar, Zeinolabedin, & Afiati, 2013).
The ITIL Core consists of five publications (Rahbar, Zeinolabedin, & Afiati, 2013):
- ❖ Service Strategy
- ❖ Service Design
- ❖ Service Transition
- ❖ Service Operation
- ❖ Continual Service Improvement

**2.3 Strategic Alignment Model**

SAM is a framework, which enables the successful implementation of business, technology and infrastructure (Henderson & Venkatraman, 1999). It identifies that business success is dependent on the concurrence of business strategy, IT strategy, organizational infrastructure and processes and IT infrastructure and processes (Rahbar, Zeinolabedin, & Afiati, 2013).

SAM is defined along two basic characteristics of strategic management (Rahbar, Zeinolabedin, & Afiati, 2013). These are: strategic fit and functional integration (Fig 1) (Rahbar, Zeinolabedin, & Afiati, 2013). Strategic fit identifies that any strategy needs to define the relationship between the internal and external domains (Rahbar, Zeinolabedin, & Afiati, 2013). The external domain (business strategy, IT strategy indicated in Fig 1) identifies how firms need to be positioned in the marketplace (Rahbar, Zeinolabedin, & Afiati, 2013). This is the part of business in which the firm competes and deals with the strategies that differentiate it from its competitors as well as making decisions about alliances and partnerships (Rahbar, Zeinolabedin, & Afiati, 2013).

The external domain addresses the following components:
1. Information technology scope focuses on those information technologies that can either support current business strategy or shape new business strategy.
2. Business scope deals with the business choices related to product-market offering.
3. Systematic competencies concentrate on those IT strategies that can create new business strategy or better support the existing one.
4. Distinctive competencies deal with those business strategies that can enable the firm to get competitive advantage.
5. IT governance focuses on the selection and use of mechanisms like strategic alliances, joint ventures etc in order to obtain the required IT competencies.
6. Business governance concentrates on 'make-versus-buy' choices including strategic alliances, joint ventures etc (Henderson & Venkatraman, 1999).
The inner area (organizational infrastructure and processes, IS infrastructure and processes as indicated in Fig 1) determines how IS infrastructure requires to be constructed and arranged (Henderson & Venkatraman, 1999).

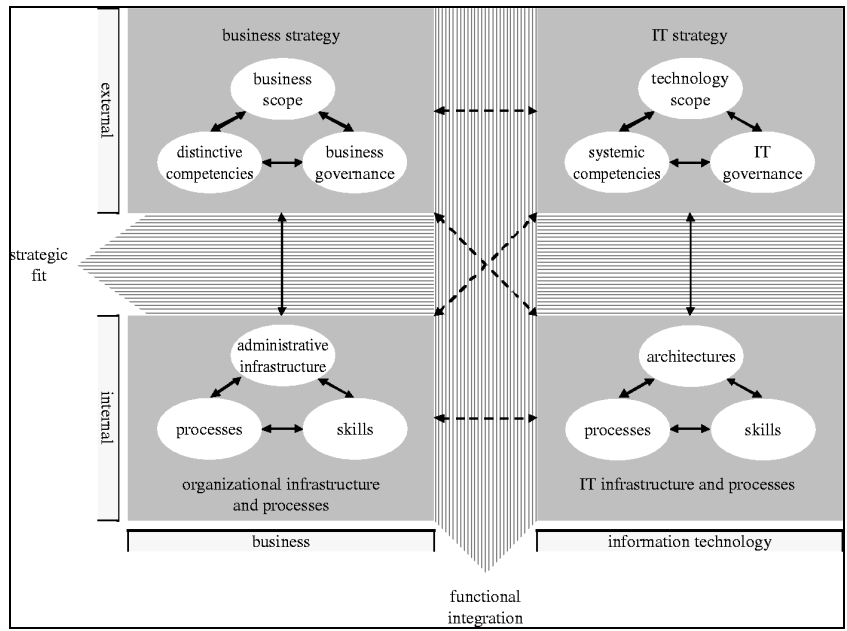

**Fig 1** Strategic Alignment Model (Henderson & Venkatraman, 1999)

### 2.4 COBIT:

CobiT is a globally accepted set of tools that executives and professionals at all organizations can use to ensure that their IT is helping them achieve their goals and objectives (ISACA, 2013). Many executives and managers need to make decisions based on diverse opinions from others, and CobiT provides a common language to better communicate goals, objectives and expected results (ISACA, 2013).

The first edition of CobiT was published in 1996 (ISACA, 2013). The second edition, in 1998, added Management Guidelines (ISACA, 2013). The third edition was released in 2000 and the fourth edition was released in December 2005, being revised and receiving the 4.1 edition in May 2007 (ISACA, 2013). CobiT 5.0 will integrate Val IT and Risk IT and is planned for 2011 (ISACA, 2013). CobiT 5.0 will also draw significantly from the Business Model for Information Security (BMIS) and the IT Assurance Framework (ITAF) (ISACA, 2013).

In addition, COBIT provides insight on how ICT processes can be launched or implemented. COBIT 4.1 has 34 high level processes that cover 210 control objectives categorized in four domains (ISACA, 2013):

Plan and Organize (PO)

- Acquire and Implement (AI)
- Deliver and Support (DS)
- Monitor and Evaluate (ME)

COBIT 5 has 37 high level processes in five domains (ISACA, 2013):

- Evaluate, Deliver and Monitor (EDM)
- Align, Plan and Organize (APO)
- Build, Acquire and Implement (BAI)
- Deliver, Service and Support (DSS)
- Monitor, Evaluate and Assess (MEA)

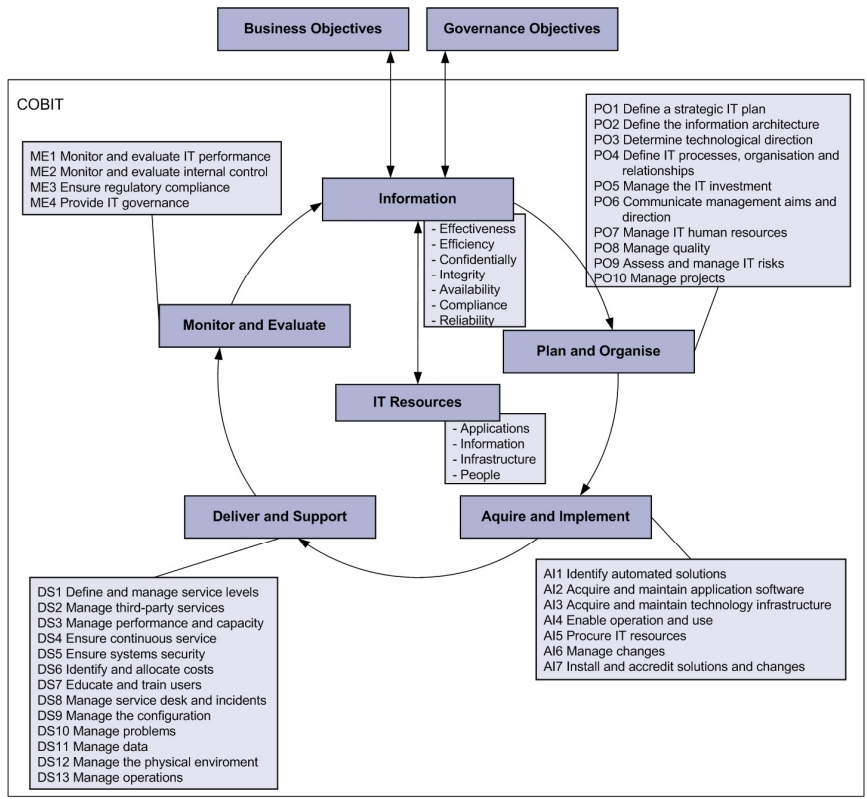

**Fig 2.** COBIT Framwork (ISACA, 2013)

## 3 RESEARCH DESIGN

This research has taken a qualitative method applying case studies. Two companies that had implemented ITIL and COBIT were chosen from two different sectors: IT and banking. One participant was selected from the IT Company and one each from the banking sector.

ITIL and COBIT have only been introduced into Iranian organizations within the last few years. Respondents were chosen based on their knowledge and experience in ITIL best practices and COBIT Framework.

## 4 FINDINGS

In the following discussion:
PA refers to: the participant from organization A, IT company;
PB refers to: the participant from organization B, the bank.

## 4.1 ITIL, COBIT and Alignment

PA determined that ITIL "Completely" is the method to attain alignment, as "there is benefit across the board".

The participant added that ITIL is the best approach and best method to improve and enhance services throughout the organization. It provides standardization and "it targets the right audiences and the right audiences are everybody – not just certain pockets". It saves money and "it communicates and streamlines processes".

ITIL improved the communication within the company and gave consistency in naming and, according to PA, "we have gone from communication of verbal to documenting the communication as well which is great". Also identified that COBIT because it is an instrument it provides users with some flexibility while giving some definite boundaries and guidelines. Additionally, it equips the user to "expand a personal approach on how to attain it and how to align". PB identified that their primary business requirement is to have service availability and reliability for their customers, ITIL assisted in delivery of those requirements.

The participator concurred that ITIL is a method to obtain alignment, as it is a framework based on people's previous experiences, problems and successes. It has a good framework and good processes in place. PB also had the same view as the others but emphasized that in order to successfully achieve alignment by COBIT the whole organization needs to embrace it. PB added "I do not think ITIL will do it by itself and I certainly do not think that IT services doing ITIL will achieve alignment. I think the whole bank has to kind of understand what it is all about and it has to be an organizational wide thing".

## 4.2 IT and Business Understanding of ITIL and COBIT

PA agreed that the business and IT people need to have an understanding of ITIL and PA stated "they should. A part of the IT, part of the strategy or business strategy should involve IT personnel and business personnel working hand-in-hand".

With regard to ITIL, also PA identified that inside the IT Company business people are one and the same and there is no gap or differentiation between them. But in general business people cannot suddenly have the same understanding and the same perspective on things in the same way IT people have.

PB identified that with regard to ITIL there is much alignment back to the business. This is because the ITIL drive is internal and external to IT and business has much knowledge of it .In the IT department they use ITIL to deliver the good services and to improve services to the business. In addition, the participant identified that from a business perspective they want to reap the benefit and "they probably do not care too much how IT is

Doing something, how our processes work or whether it is ITIL based or whatever, as long as they are getting the service that they require". Additionally, PB continues some of the IT people are so technical but some of the ITIL processes such as change process, service asset process, event and incident response time, and service level agreements requirements pressure them to take the business face and the effect on the business into account.

# 5 DISCUSSIONS

The meaning of this research was to try to find out whether or not alignment could be attained by introducing ITIL and COBIT. The effect of ITIL and Cobit and Also affect on each of these views will be analyzed in this part.

Taking into account the communication between business and IT in the organizations under study, the initiatives they have taken to make improvements, plus the fact that ITIL and COBIT can improve communication, it would be interesting to find out if it is possible to align business and IT by applying SAM and the four dominant alignment perspectives of it (N.Zeinolabedin and etal,2013).These perspectives identify how to make an alignment between the business strategy, IT strategy, organizational infrastructure and processes and IT infrastructure and ITIL and COBIT processes (Fig 3). In order to understand the alignment Perspectives in detail IT infrastructure, ITIL and COBIT processes quadrant will be discussed in the next section. In addition the impact of this quadrant on the other three parts also has been determined. If the conclusion is reached that ITIL and COBIT have the ability to make alignment between these four perspectives, then the research question will be answered. These alignment perspectives are as follow.

## 5.1 Business Strategy Execution, ITIL and COBIT

In this perspective the manager's role is to create and formulate business strategy and for IT managers to implement those strategies.

IT needs to review business requirements and ensure they can deliver those requirements in terms of time, cost and resources.

They use ITIL and COBIT to provide some direction for the business and to improve their strategies. For organization A the participants identified that they are an IT company and there is no differentiation between IT people and business people so they are write only one business strategy. Therefore, in this organization IT supports and develops business strategy. But The IT people in organizations B were not so involved in developing the business strategy only in delivering them.

In order to understand this alignment perspective better, the impact of ITIL and COBIT on each of the other two quadrants (business strategies; organizational infrastructure and processes) will be discussed. But firstly we discuss the IT infrastructure and ITIL processes quadrant.

**IT Infrastructure and ITIL Processes (Fig 3):**

This quadrant included three processes: IT infrastructure, ITIL processes and ITIL skills. ITIL processes are significant to the operation of IT infrastructure. The main processes are service transition and service operation, which enable organizations to deliver services more efficiently and effectively.

The IT infrastructure required for ITIL to work has been identified from different perspectives by the participants. One identified that IT infrastructure needs to be structured centrally. Another identified that for ITIL to work tools, technologies and buy-in at every level are required. But the other two stated that ITIL can work in any infrastructure because it is the operation of services that is important not just the kind of infrastructure that organizations have. However, it has been identified that the impact of ITIL on the operation of IT infrastructure has been significant as it is a framework and provides guidelines for hardware and software. The findings identified that areas mostly affected by ITIL processes are service operation and service transition functions.

It has also been identified that for ITIL to be successfully operated in organizations some skills and training are required. The participants identified they are providing training for only some of their staff because of the cost but they have an adequate number of experts to work with ITIL.

**Business Strategy (Fig 3):**

With regard to the ability of ITIL and COBIT in supporting and shaping business strategy the findings of this research have identified that ITIL and COBIT can support and shape the business strategy. It enables organizations to differentiate themselves from their competitors by improving the quality of services they deliver to customers. Also the ITIL and COBIT processes improve delivery of services, the quality of those services and at low cost to organizations.

So ITIL and COBIT impacts on each of the processes inside the business strategy quadrant. These processes are business scope, distinctive competency and business governance.

## Organizational infrastructure and processes (Fig 3):

There was disagreement by participants on the effects of ITIL and COBIT on business processes. The majority of participants identified that the impact of ITIL on business processes was minimal, in the other hand they stated that the business processes were heavily impacted by COBIT. In addition, the findings identified that COBIT impacts on communication inside organizations breaks down barriers and improves knowledge sharing. As a result, the skills required to execute business strategies are enhanced. Moreover, the findings revealed that ITIL and COBIT processes can heavily impact the operation of IT infrastructure.

Therefore ITIL and COBIT can impact on (business process, business skills and administrative infrastructure). These are the processes inside this quadrant.

Overall, for this alignment perspective, it can be concluded that for two of the organizations studied IT supports business strategy and shapes business strategy because of the nature of the company. In terms of ITIL and COBIT, the two organizations agreed that ITIL and COBIT supports business strategy by their processes. This alignment perspective has been approved as business strategy formulated by business managers can be supported and implemented by IT using ITIL (N.Rahbar and etal,2013)and COBIT.

### 5.2 Technology Transformation , ITIL and COBIT

This perspective focuses on the implementation of business strategy through appropriate IT strategy and IT infrastructure and ITIL processes. The extent to which business strategy is implemented through IT strategy and the required IT infrastructure and ITIL processes has been identified by this research. The findings from organizations A and B identified that the business strategy can be implemented by ITIL (N.Rahbar and etal,2013) and COBIT processes but in an indirect way – it depends on the business strategy and where it needs to be reflected.

**IT Strategy (Fig 3)**:

The findings of the research have identified that ITIL and COBIT have the ability to impact on IT strategy and its processes (technology scope, IT governance and IT competency).

ITIL and COBIT impacts positively on IT competency and can make IT more efficient and effective in delivering business needs, strategy and strategic direction. It enables organizations to utilize their IT infrastructure and deliver better services at less cost. Maintaining the cost means focusing more on business strategy rather than trying to fix things up all the time. ITIL and COBIT enable organizations to view the "bigger picture" rather than just focusing on a new technology or specific project. They also enable organizations to grow their technology scope by focusing on business strategy and attempting to deliver to that. Even though the impacts on these two parts were significant, there were disagreements among organizations about its impact on IT governance.

Two of the organizations identified that the impact on their IT governance was "phenomenal" and it enabled them to provide better technology and solutions for their customers.

Business strategy can be implemented by ITIL(N.Rahbar and etal,2013) and COBIT processes indirectly. ITIL and COBIT have the ability to impact significantly on both the business strategy and IT strategy. This alignment perspective has also been approved.

### 5.3 Competitive Potential , ITIL and COBIT

This perspective enables the adoption of business strategy through IT capabilities. The previous section discusses how ITIL and COBIT can impact on business strategy – they can support and shape it. In addition, it has been identified that ITIL and COBIT impacts heavily on IT strategy and improves it significantly.

ITIL and COBIT can impact on organizational infrastructure significantly. In addition, it was discovered that ITIL and COBIT improves knowledge sharing and communication inside organizations. Therefore, it enables organizations to better execute business strategy. This alignment perspective has also been approved (N.Rahbar and etal,2013).

### 5.4 Service Level, ITIL and COBIT

This perspective concentrates on providing the best IT service for organizations by focusing on IT strategy and the IT infrastructure and ITIL processes. As has been discussed in the previous sections, ITIL and COBIT can impact IT strategy by improving the IT competency through enabling organizations to utilize the IT infrastructure and provide good services at low cost. Also, they impact on technology scope by better supporting business strategy. The impact on IT governance however, has been identified as being different for organizations.

It has also been identified that they impact organizational infrastructure and processes and can improve them. This means that IT has the ability to better deliver services and at low cost. The findings identified that the benefits the ITIL and COBIT provided for organizations were mainly improving delivery of services, internal communication, identifying the areas of weakness and having consistency for the way things need to be done in organizations. All of these increased the ability of organizations to have better delivery of services now and increased delivery of services in the future. This alignment perspective has also been approved.

## 6 CONCLUSIONS

This research attempted to resolve the issue of alignment by investigating the impact of ITIL and COBIT, which are a set of best practices enabling organizations to deliver their services more efficiently, effectively and at less cost. The research intended to identify how effective ITIL is in making alignment between IT and business objectives.

Employing ITIL to the SAM model has ascertained that ITIL can affect on the four alignment perspectives introduced by the model. It has been recognized that ITIL has the potential to influence business strategy, support and transform it. More ever, it can alter and magnify IT strategy. In addition the COBIT is an instrument that tries to solve common project problems, in an integral, consistent document, and a process of continuous improvement. Much of COBIT is unique to project management, such as critical path and work breakdown structure (WBS). Some areas overlap with other management disciplines. IT must add value to an organization; it must develop its competence in areas such as teamwork, negotiation, resolution of conflicts, communication, risks and in the definition of effective processes. These competences are presented in the COBIT in the form of a body of knowledge that unifies processes, Knowledge, necessary skills and techniques for the management of projects.

The outcomes of this investigative research specified that ITIL and COBIT are successful approaches in creation alignment in organizations.

### 6.1 Implication for Practice and Limitations and Future Study

It has been recognized formerly that attaining strategic alignment is the number one issue faced by management. This research introduced ITIL and COBIT as approaches to achieve alignment and the findings confirmed that they have the ability to do so. Additionally, ITIL and COBIT can be used together to address the three primary aspects of IT governance:1) conformance 2) performance 3) relating responsibility. Therefore, using ITIL and COBIT as a way of achieving alignment could be a successful strategy for management.

The initial restriction of this study was that because there were a few number of organizations deliberated, reliability of the announcements is the least. To make the research more reliable a number of other organizations need to be investigated. One more restriction was that as applying ITIL and COBIT to achieve strategic alignment are a new idea to Iranian organizations, limited research has been done in this zone, which makes it arduous to contrast the outcomes of this study with others. With regard to the possible for next research, it has been recognized from this research that ITIL and COBIT have the potential and are effectual approaches to make alignment between IT and business targets.Measuring the extent to which ITIL and COBIT can make alignment could be the topic of further research.

Fig3 ITIL and COBIT Strategic Alignment Model

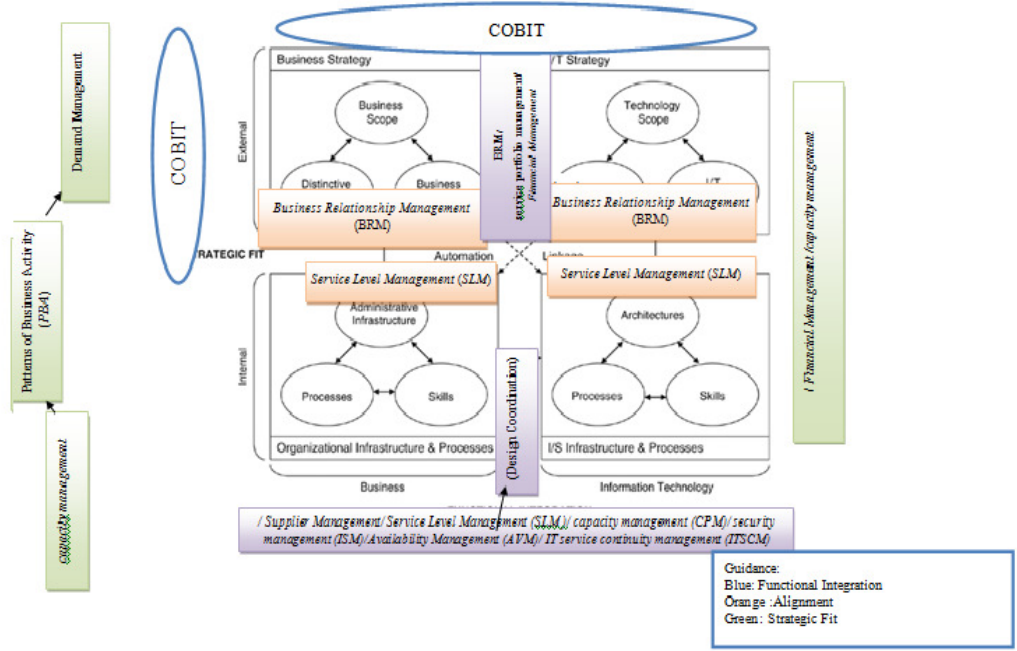